\documentstyle[12pt,aasms4]{article}  

\def\hi{H~{\small I}}
\def\hii{H~{\small II}}  

\begin{document}

\noindent  

{\it Conference Highlights}

\begin{center}  


\title{\large \bf Galaxy Disks and Disk Galaxies\footnote{Conference
was held in Rome, Italy at the Pontifical Gregorian University on June
12-16, 2000. Proceedings will be edited by J.G.\ Funes, S.J. \& E.M.\
Corsini and published in the {\it ASP Conference Series}}}
  
\end{center}

\medskip  


The conference Galaxy Disks and Disk Galaxies, sponsored by the
Vatican Observatory, was held in June 2000 at the Pontifical Gregorian
University, in the very center of Rome.  The meeting hosted about 230
participants coming from 30 countries.  Social events included a visit
of the Vatican Observatory and the Papal Villas at Castel Gandolfo. A
buffet supper was served on one of the terraces of the Papal Summer
Residence where the Vatican Observatory is located.  The very full
program consisted of 29 review papers, 34 invited talks, and more than
180 posters. The meeting covered topics regarding the structure,
formation and evolution of galaxies with disks. Particular attention
was dedicated to the stellar and gaseous disk of the Milky Way, the
global characteristics of galaxy disks, their structure, morphology
and dynamics, the gaseous components, star formation, and chemical
evolution, the interactions, accretion, mergers and starbursts, the
dark and luminous matter, the establishment of the scaling laws,
and the formation and evolution of disk galaxies from a theoretical
and observational point of view.

New observational evidences constraining the star formation and  
merging history of the Galaxy were presented. There was a broad  
consensus about the mass distribution of the Milky Way. NIR data show  
that the Galactic bar extends to about 4 kpc while the terminal curve,  
the Oort limit, and the bulge microlensing observations together  
favour the idea of a massive and near-maximal disk.  The vertical  
structure the Galactic disk, and in particular the origins of heating,  
was explored both from the empirical and theoretical point of  
view.  A new composite CO survey provided a nearly complete inventory  
of giant molecular clouds, which are one of the main sources for disk  
heating. The correlation between the thickening of the Galactic disk,  
its stability, and the fraction of the total mass to be attributed to  
the Galactic halo have been taken into account and tidal tails from  
satellites have been used to  probe the gravitational potential. Chemical  
abundances and abundance ratios derived for the Galactic disk impose  
strong constraints on its formation and evolution and they indicate an  
inside-out process of growth which is still ongoing.   

Analysis of the surface photometry of a sample of around 750
early-type galaxies reveals that elliptical galaxies also have disks,
suggesting that all galaxies were either born with, or later acquired,
disks. The scaling relations of typical disks in ellipticals smoothly
join those of spiral and S0 disks, even if some marked differences do
exist for disks deeply embedded in spheroids. New $K-$band images show
that the morphological structure of disk galaxies differs from that
observed at shorter wavelengths, in particular the decoupling between
the stellar and the gaseous disks can be dramatic.  In any event, the
co-existence of radically different morphologies within the same
galaxy can be explained naturally by the modal theory of spiral
structure. Large-scale magnetic fields, remarkable in their extent and
close correlation with the spiral arms, have been detected in the
disks of all nearby galaxies. Sometimes they show vertical orientation
suggesting field closure through the halo. Their origins have been
interpreted in terms of a multi-mode dynamo theory.
  
Theoretical efforts that take into account the dynamical processes in
self-gravitating disks have been made in order to explain the large
spiral structure observed in the optical and the near-infrared. The
role of spiral arms and bars in driving the evolution of disks was
also discussed, showing that we now have a mature theory of disk
instability. The secular evolution of disks proceeds through
self-regulation and feedback mechanisms. Dynamical modeling has 
benefited greatly from the unique capabilities of integral field
spectroscopy. Observation of external galaxies suggests that the
dominant heating mechanism varies along the Hubble sequence and has
shown the existence of cyclones and anticyclones in gaseous disks.  An
update of the correlation between black-hole mass and the mass of the
host spheroidal component was presented (the black-hole mass fraction
is $\sim0.17\%$). It also appears that the black-hole mass is even
more strongly correlated with the stellar velocity dispersion of the
host galaxy.  This new result corroborates the idea that black holes,
quasars and bulges grew and turned on as parts of the same process.

Neutral hydrogen in and around some 200 spiral and irregular galaxies
has now been mapped with the WHISP survey. The \hi\ disks are more
extended than the corresponding optical disks and show a wider variety
of features such as debris of tidal interactions, warps, lopsidedness,
spiral arms and even bar-shaped structures.  Although \hi\ structures
are rare in giant elliptical galaxies, they appear to be more common
in low-luminosity systems. Their morphology ranges from chaotic,
through loops and rings, to still-growing disks with surprisingly
regular kinematics.  New interferometric surveys have enhanced our
knowledge of the distribution and kinematics of molecular gas on
sub-kiloparsec scales, and impressive {\it Chandra\/} pictures
unveiled the violent side of the ISM by mapping the hot gaseous
component in nearby galaxies.  Characterizing the demographics and
quantifying the spatial variation of the star forming galaxy
population, as well as probing the dependence of the local SFR on the
properties and dynamics of the ISM, turned out to be hot topics for a
detailed understanding of star formation in galaxy disks. Their most
luminous \hii\ regions seems to be density bounded and responsible for
half of the H$\alpha$ emission in normal disks.

Great emphasis was given to the role played by galaxy interactions and
mergers in the nearby universe, and even more so in the early
universe, where galaxies appear to be far from equilibrium. 
Numerical simulations are widely used to explore the merging
processes driving morphological transformations, and show that
low-luminosity fast-rotating ellipticals are not formed as byproducts
of collisionless mergers of unequal-mass disk galaxies. Beyond this, a
complete picture of the role and the fate of gas has not yet been
produced and it deserves further investigation.  Extended disks form
by accreting material from the environment and are often warped, most
probably due to a continuously maintained misalignment between the
angular momenta of the galaxy and its halo. A large fraction of
spirals exhibit kinematic disturbances ranging from mild to major, and
can generally be explained as the visible signs of tidal
encounters. Planetary nebulae around ellipticals can be used to trace
the gravitational potential and interactions with the surrounding
cluster. Gravitational interactions are also likely to be the main mechanism
for both the transformation of cluster spirals into S0s and of the
fuelling of gas into the center of the encounter partners, thus
triggering star formation.

New combined optical and radio rotation curves were presented both for
high and low surface brightness spirals. They were used to derive mass
models and constrain the size of dark halos. LSB and dwarf galaxies
are dominated by dark matter in agreement with early
results based on \hi\ data alone. Closer to us, the kinematics of the
LMC is consistent with a truncated, finite-thickness exponential disk
model with no dark halo. The origin of scaling laws related to the
structural properties of disk galaxies were discussed with the help of
high-resolution gas-dynamical simulations and semi-analytic models, and
a lively debate arose on the controversy about the relative fraction
of luminous to dark matter inside the optical region of
spirals. Observational evidence, modeling results, and 
theoretical considerations were provided for and against the idea that
stellar disks contribute a maximal rotational support in the inner
parts of galaxies. Two of the most promising developments presented
were the discovery of rapidly rotating bars through the direct
measurement of their pattern speed and the better agreement
between the predicted and observed Tully-Fisher zero point obtained by
semi-analytical modelers through a proper choice of the initial power
spectrum.

General agreement on the scheme for galaxy formation based on
hierarchical clustering and mergers was the the strongest point of
convergence of the meeting. The formation of bulged galaxies is
a problem which has been essentially resolved. Large
bulges are assembled by mergers of subclumps, formed from gas cooled
in the center of dark halos, and then acquire their disks, while small
bulges are most probably the result of disk secular evolution. The
formation of pure-disk galaxies still remains an open question which
must be addressed. New ideas have focussed on the formation and
structure of dark matter halos, including their density and
angular-momentum profiles, abundance, and merging rates. Evidence
supporting the idea that damped Ly$\alpha$ systems observed at
$z\sim3$ are the direct progenitors of the current thick disks in
galaxies was discussed, and there were indications that some extremely
red objects at high redshift are disks. The chemical history of Local
Group galaxies has been used to infer the nature and evolutionary
status of higher-redshift objects and new spectrophotometric codes are
now available to follow galaxy evolution and to constrain cosmic star
formation history.

As pointed out in the Conference Summary, galaxies are the cross roads
of Astronomy because they look up to Cosmology and they look down to
the interstellar medium and star formation. This meeting 
highlighted the points of convergence in our comprehension of the
formation and properties of disk galaxies, suggesting at the same time
that the future challenges in this field are to make those connections
(such as between the properties of hydrodynamical models and those of
real galaxies or between a physically based theory of star formation
and what is observed in galaxies) that at the moment we have failed to
make.


\noindent  
{\it Jos\'e G. Funes, S.J. and Enrico Maria Corsini}   
 
\noindent  
{\it Vatican Observatory, Universit\`a di Padova}  

\end{document}